\documentclass[aps,prl,twocolumn,superscriptaddress,floatfix,nofootinbib,showpacs,longbibliography]{revtex4-2}

\usepackage[utf8]{inputenc}  
\usepackage[T1]{fontenc}     
\usepackage[british]{babel}  
\usepackage[table]{xcolor}
\usepackage[scaled=0.86]{berasans}  
\usepackage[colorlinks=true, citecolor=blue, urlcolor=blue]{hyperref}
\usepackage{comment}
\usepackage{bm}
\makeatletter
\newcommand{\setword}[2]{%
  \phantomsection
  #1%
  \protected@edef\@currentlabel{#1}%
  \label{#2}%
}
\makeatother
\makeatother
\usepackage{comment}
\usepackage{graphicx} 
\usepackage[babel]{microtype}  

\usepackage{amsfonts}
\usepackage{amsmath}
\usepackage{amssymb,amsthm}

\usepackage{xspace}  
\usepackage{pgf,tikz}
\usepackage{xcolor}
\usepackage{multirow}
\usepackage{array}
\usepackage{bigstrut}
\usepackage{physics}
\usepackage{color}
\usepackage{natbib}
\usepackage{multirow}
\usepackage{mathtools}
\usepackage{dsfont}
\usepackage{xcolor,colortbl}
\usepackage{color}

\newcommand{\be}{\begin{equation}}
\newcommand{\ee}{\end{equation}}
\newcommand{\ba}{\begin{eqnarray}}
\newcommand{\ea}{\end{eqnarray}}

\newcommand{\Span}{{\mathsf{Span}}}

\def\>{\rangle}
\def\<{\langle}

\usepackage{centernot}
\usepackage[caption=false]{subfig}

\providecommand{\ket}[1]{| #1{\rangle}}
\providecommand{\bra}[1]{\langle #1|}

\usepackage{xcolor}

\usepackage{newtx}

\begin{document}

\title{Gaussian superpositions for bosonic encodings}

\author{Federico Centrone}
\email{fcentrone@icfo.net}
\affiliation{ICFO-Institut de Ciencies Fotoniques, The Barcelona Institute of Science and Technology, Av. Carl Friedrich Gauss 3, 08860 Castelldefels (Barcelona), Spain}
\affiliation{Universidad de Buenos Aires, Facultad de Ciencias Exactas y Naturales,
Departamento de Física, Ciudad Universitaria, 1428 Buenos Aires, Argentina}

\author{Juan Pablo Paz}
   \affiliation{Universidad de Buenos Aires, Facultad de Ciencias Exactas y Naturales,
Departamento de Física, Ciudad Universitaria, 1428 Buenos Aires, Argentina}
    \affiliation{CONICET - Universidad de Buenos Aires, Instituto de Física de Buenos Aires (IFIBA),
Ciudad Universitaria, 1428 Buenos Aires, Argentina}

\author{Augusto Roncaglia}
    \affiliation{Universidad de Buenos Aires, Facultad de Ciencias Exactas y Naturales,
Departamento de Física, Ciudad Universitaria, 1428 Buenos Aires, Argentina}
    \affiliation{CONICET - Universidad de Buenos Aires, Instituto de Física de Buenos Aires (IFIBA),
Ciudad Universitaria, 1428 Buenos Aires, Argentina}

\date{\today}
    \begin{abstract}
Quantum harmonic oscillators and Gaussian states are among the elementary building blocks of physics, describing linear waves, bosonic modes, and free quantum fields, whereas qubits are the elementary carriers of quantum information. Finite superpositions and mixtures of Gaussian branches provide a natural bridge between these two languages: they are the simplest non-Gaussian extension of Gaussian physics, with Schrödinger cat states as the paradigmatic example, and they arise across quantum optics, superconducting circuits, and relativistic detector-field models, with error correction being one of the main applications. However, bosonic Hilbert spaces are infinite-dimensional and, beyond the Gaussian regime, the spectral properties required for quantum-information diagnostics such as entropy and entanglement become prohibitive in multimode or high-energy regimes.
We show that any density operator supported on a finite span of pure Gaussian branches admits an exact finite-dimensional logical density matrix whose nonzero spectrum coincides with that of the original infinite-dimensional bosonic operator. This makes mixed-state entropies and encoded entanglement accessible directly from Gaussian overlaps and cross moments, even in highly excited multimode regimes where direct Fock-space diagonalization is impractical; the finite spectral problem scales with the number of branches rather than a photon-number cutoff. We illustrate the construction by calculating mixed-state relative-entropy non-Gaussianity and the negativity of multimode two-branch product superpositions. The algebraic reduction applies more generally to finite nonorthogonal families with computable overlaps.
\end{abstract}

\maketitle

\paragraph{Introduction.}
Quantum information theory is traditionally formulated in finite-dimensional Hilbert spaces, and its flagship tools, e.g., entropies, distinguishability measures, and entanglement monotones such as the negativity, are designed for qubits and qudits. Yet many leading platforms for quantum technologies and fundamental tests, like quantum optics, cavity and circuit QED, optomechanics, and relativistic quantum field theory, are built from bosonic modes with intrinsically infinite-dimensional state spaces \cite{Brune1996Cat,Haroche2013RMP}. In these settings, the most tractable operations and states are Gaussian, but Gaussian resources alone are insufficient for universal quantum computation and many tasks that require genuinely nonclassical features \cite{walschaers2021non}. The relevant operating point is therefore a hybrid one: dynamics and control are often largely Gaussian, while non-Gaussianity is injected by measurements, nonlinearities, coupling to matter, or conditioning, producing states that are hard to manipulate \cite{Vlastakis2013Cat, leghtas2015confining}.

The natural non-Gaussian structure generated in this hybrid regime is often a finite superposition (or mixture) of Gaussian branches \cite{braccini2025superpositions}. Canonical examples are Schr\"odinger-cat states (superpositions of coherent states), which have evolved from early cavity-QED demonstrations into a versatile bosonic toolbox for control, sensing, and error correction \cite{Brune1996Cat,leghtas2015confining,ofek2016extending, PazRevoraSchmiegelow2025}
. More broadly, Gaussian branch structures arise whenever a bosonic mode becomes correlated with a discrete record, such as an ancilla qubit, a detector click, or an unobserved measurement trajectory, so that each conditional branch remains Gaussian while the overall state is non-Gaussian. Related branch interference also appears in relativistic settings with quantum-controlled couplings and non-equilibrium dynamics, where distinct dynamical histories contribute coherently \cite{foo2021thermality, FooOnoeZych2020_UDWsuperpositionTrajectories,BarbadoCastroRuizApadulaBrukner2020}. Importantly for quantum information, many bosonic \emph{encodings} are naturally expressed in this language: cats are two-branch encodings, symmetry-generated ``multiplets'' encode qudits in a single oscillator, and finite-squeezing approximations of toroidal grid/GKP-type codewords can be viewed as superpositions of many squeezed displaced Gaussian wavepackets \cite{GottesmanKitaevPreskill2001,Michael2016GKP,CampagneIbarcq2020GKP,Grimsmo2020MultiphotonCodes}. Across these settings,  Gaussian noise is a central practical limitation \cite{SerafiniParisAdesso2004,Mirrahimi2014NJP_catQubits,ofek2016extending}.  Recent work has shown that Gaussian superpositions are not merely a compact ansatz but, in several settings, the representation that makes otherwise inaccessible problems tractable. These methods enable efficient simulations of cat, GKP, and related bosonic qubits under Gaussian channels, measurements, and non-Gaussian transformations~\cite{bourassa2021fast,marshall2023simulation,dias2024classical,hahn2025classical,solodovnikova2025fast,braccini2025superpositions}. In relativistic quantum field settings, exact control of the relative phase between Gaussian branches has been used to model quadratic-detector communication in a regime previously inaccessible to standard methods~\cite{funai2025gaussian}. Earlier analyses of nonorthogonal entanglement and coherent-state encodings already exploit finite logical descriptions~\cite{wang2002nonorthogonal,vanEnkHirota2001,JeongKim2002,vanEnk2005,Sanders2012}. These developments motivate a systematic spectral treatment extending beyond these special cases to arbitrary pure Gaussian branches, mixed states, and information-theoretic quantities requiring both Gaussian overlaps and cross moments.

This motivates a question that is both conceptual and practical: \emph{once a non-Gaussian bosonic state is generated or represented as a finite Gaussian-branch state, can its spectral quantum-information content be extracted without introducing a local Hilbert-space cutoff?} In this work, we answer yes. We construct an exact finite-dimensional density matrix for any state, pure or mixed, whose support lies in the span of a finite, generally non-orthogonal set of pure Gaussian states. This finite matrix preserves the nonzero spectrum of the original infinite-dimensional density operator and therefore supports the direct use of standard discrete-variable tools, such as spectral entropies, distinguishability measures, purity, and, when the branch manifold factorizes across a bipartition, PPT/negativity tests. All dependence on the underlying continuous-variable structure enters through Gaussian overlaps and, when needed, low-order cross moments that are available in closed form from Gaussian data. 

This perspective yields an analytic and numerical way to quantify highly excited, multimode, non-Gaussian, and mixed bosonic states in regimes where Fock-space truncations become unreliable or infeasible. In particular, it turns Gaussian-branch representations, previously used mainly for propagation, sampling, or expectation values, into exact finite spectral objects. We use the same finite density matrix to obtain cutoff-free relative entropy non-Gaussianity for mixed two-branch states and a closed expression for the bipartite negativity of arbitrary multimode two-branch product Gaussian superpositions, where the result depends only on the overlaps of the Gaussian states.

\paragraph{Logical density matrix.}
\label{par:finite_manifold}

We now construct this finite-dimensional representation explicitly. Once this matrix is constructed, one can apply standard discrete-variable quantum information tools directly to the encoded manifold, without introducing a photon-number cutoff. The construction uses the Gram matrix of the non-orthogonal Gaussian branches; L\"owdin symmetric orthogonalization~\cite{lowdin1950non} provides a convenient orthonormal procedure, but all spectral quantities are independent of this choice.

Let $\mathcal H$ be a bosonic Hilbert space and let
$\mathcal B_D=\{\ket{g_k}\}_{k=0}^{D-1}\subset\mathcal H$ be a finite set of
normalized, linearly independent pure Gaussian states. The states $\ket{g_k}$ need not be
orthogonal, and their span
\begin{equation}
\mathcal M_D:=\Span(\mathcal B_D)
\end{equation}
generally contains non-Gaussian states. The only Hilbert-space data needed to construct
the finite logical representation are the branch overlaps
\begin{equation}
G_{jk}:=\braket{g_j}{g_k},
\qquad
G=G^\dagger,\qquad
G\succ0,
\label{eq:Gram_main}
\end{equation}
where $G$ is the Gram matrix, whose positivity follows from linear independence. For pure Gaussian states, these
overlaps are closed-form Gaussian integrals.

To obtain an orthonormal basis of $\mathcal M_D$, collect the branches into the
formal column
\begin{equation}
\Psi:=\big[\,\ket{g_0}\ \cdots\ \ket{g_{D-1}}\,\big],
\qquad
\Psi^\dagger\Psi=G .
\label{eq:Psi_main}
\end{equation}
L\"owdin orthogonalization defines
\begin{equation}
\Phi:=\Psi G^{-1/2}
=
\big[\,\ket{s_0}\ \cdots\ \ket{s_{D-1}}\,\big],
\label{eq:Phi_main}
\end{equation}
so that $\Phi^\dagger\Phi=I_D$. The states $\ket{s_k}$ are not, in general,
Gaussian states; they are an orthonormal logical basis for the Gaussian-branch
manifold. A different orthonormal basis would only conjugate the finite matrices by a
unitary, leaving spectra and trace functions unchanged.

Now consider any density operator supported on $\mathcal M_D$, which can be written as
\begin{equation}
\rho=\sum_{i,j=0}^{D-1}c_{ij}\ketbra{g_i}{g_j}=\Psi \, {\mathbf C} \, \Psi^\dagger 
\label{eq:mixed_branch_state_main}
\end{equation}
with $\mathbf C$ a $D\times D$ hermitian matrix such that ${\rm tr}( {\mathbf C}G)=1$ and $G^{1/2}\mathbf C G^{1/2}\succeq0$. In the orthonormal basis, the state is represented as:
\begin{equation}
\rho=\Phi \, (G^{1/2}{\mathbf C}G^{1/2}) \, \Phi^\dagger 
\label{eq:mixed_branch_state_main2}
\end{equation}
 On the other hand, pure states can also be represented as:
\begin{equation}
\ket{\psi(\mathbf c)}=\sum_{k=0}^{D-1}c_k\ket{g_k}=\Psi \mathbf c,
\qquad
\braket{\psi(\mathbf c)}{\psi(\mathbf c)}=\mathbf c^\dagger G \mathbf c =1
\label{eq:pure_branch_state_main}
\end{equation}
with $\mathbf c$ a complex vector satisfying the above normalization condition.  In the orthonormal basis, the pure states are represented as: $\ket{\psi(\mathbf c)} = \Phi G^{1/2}\mathbf c$. Notice also that the $\mathbf C$ matrices associated with pure states are such that $\mathbf C = \mathbf c\,  \mathbf c^\dagger.$

The finite density matrix on the branch support is not only a matrix of branch amplitudes: it has the same nonzero spectrum as the original bosonic operator. 
Thus von Neumann and R\'enyi entropies, trace norms, distinguishability measures,
hypothesis-testing quantities, and entanglement
tests reduce to ordinary finite-dimensional linear algebra on this exact density
matrix. This representation applies
to arbitrary density operators supported on \(\mathcal M_D\), including states with
coherences between different Gaussian branches.

The spectral reduction itself is algebraic and applies to any finite linearly independent branch family with known overlaps. The useful point here is that, in the bosonic setting considered in this work, the finite support is specified by Gaussian branches: the entries of the Gram matrix, and the cross moments needed for phase-space diagnostics, are available as closed-form Gaussian integrals. This makes the construction operational for highly excited and multimode non-Gaussian states, precisely where Fock-basis truncations become costly or ill-conditioned.
Related Gram-matrix and
non-orthogonal-basis methods have been developed for quantum metrology
\cite{genoni2019non,fiderer2021general}; here the same finite-support structure is used
to build an exact density matrix for Gaussian-branch bosonic states, from which general
quantum-information quantities can be computed.

A common special case occurs when the branches form a finite orbit
$\ket{g_k}=U^k\ket{g_0}$, with $U^D=I$. Then
$G_{jk}=\bra{g_0}U^{k-j}\ket{g_0},
$ so $G$ is circulant and is diagonalized by the discrete Fourier transform,
$G=F^\dagger\Lambda F, \:
\Lambda=\mathrm{diag}(\lambda_0,\ldots,\lambda_{D-1})$, 
with
$\lambda_m
=
\sum_{k=0}^{D-1}
\bra{g_0}U^k\ket{g_0}\,
e^{-2\pi i m k/D}$.
This includes rotation-symmetric bosonic encodings, where $U$ is a finite phase-space rotation. Finite Gaussian approximations to grid states also fit the general construction, but need not arise from an exact finite-order displacement orbit.

The representation above is exact for states whose support lies in the chosen branch
manifold. Generic noise channels need not preserve this finite support, unless the
branch set is enlarged or the channel has a compatible finite-branch action. In this sense, finite Gaussian-branch manifolds define a tractable non-Gaussian sector: they are not Gaussian and need not be low-energy, yet their quantum-information content can be extracted from finite matrices built from Gaussian data.

The next two sections use the same exact density matrix to treat two quantum-information diagnostics: relative entropy non-Gaussianity, which also requires Gaussian cross moments to construct the reference Gaussian state, and bipartite negativity, which, for two-branch encoded Bell states, is determined by the finite representation and Gaussian branch overlaps. A more detailed derivation of the finite-manifold reduction is given in
\hyperref[app:finite_manifold]{Appendix A}.

\paragraph{Entropy of non-Gaussian states.}
We first use this representation to answer a basic question: how much non-Gaussianity is carried by a mixed Gaussian-branch state? The state is supported on two
pure Gaussian branches $\{\ket{g_1},\ket{g_2}\}$, 
\begin{equation}
\ket{\psi_\pm}
=\frac{1}{\sqrt{\mathcal Z_\pm}}\Big(\ket{g_1}\pm \kappa\,\ket{g_2}\Big), \:
\mathcal Z_\pm=1+\kappa^2\pm 2\kappa\,\Re\!\braket{g_1}{g_2},
\label{eq:two_branch_state_pm}
\end{equation}
where $\kappa\ge 0$ is a real amplitude ratio (we fix the relative phase so that $\braket{g_1}{g_2}\ge 0$).
To access mixed states while preserving the two-dimensional support, we model preparation uncertainty between the two relative phases $0$ and $\pi$:
\begin{equation}
\rho_{p}
=
(1-p)\ket{\psi_+}\bra{\psi_+}
+
p\ket{\psi_-}\bra{\psi_-},
\qquad 0\le p\le 1.
\label{eq:phase_randomization_model}
\end{equation}
For well-separated or strongly squeezed branches, these states contain many
photons, so a Fock-basis calculation of $S(\rho_p)$ needs a large cutoff,
whereas the representation above is exact.

We quantify non-Gaussianity by the relative entropy to the Gaussian reference state $\tau(\rho)$ with the
same first and second moments:
\begin{equation}
\delta_{\rm nG}(\rho):=S\!\big(\rho\|\tau(\rho)\big)
= S\!\big(\tau(\rho)\big)-S(\rho),
\label{eq:delta_nG_def_main}
\end{equation}
where the identity follows because $\tau(\rho)$ maximizes the von Neumann entropy among all states with the same
covariance matrix \cite{WolfGiedkeCirac2006,Genoni2008PRA_relativeEntropyNG,GenoniParis2010}. Operationally, $\delta_{\rm nG}(\rho)$ quantifies the asymptotic distinguishability of $\rho$ from its closest Gaussian state. Indeed, the reference state $\tau(\rho)$, determined by the first and second moments, minimizes $S(\rho\Vert\sigma_G)$ over all Gaussian states $\sigma_G$~\cite{marian2013relative}. In asymmetric quantum hypothesis testing between independent copies of $\rho$ and $\tau(\rho)$, the quantum Stein lemma identifies $\delta_{\rm nG}(\rho)$ as the optimal asymptotic error exponent for mistaking the Gaussian reference for $\rho$, while keeping the opposite error probability bounded~\cite{ogawa2000strong}. Thus, for a sufficiently small target error $\epsilon$, the required number of copies scales asymptotically as $\ln(1/\epsilon)/\delta_{\rm nG}(\rho)$, assuming optimal collective measurements. This interpretation applies also to mixed states. Its evaluation requires the von Neumann entropy of the non-Gaussian state, which we obtain directly from the finite logical spectrum.

Within our framework, both terms in \eqref{eq:delta_nG_def_main} are obtained without any photon number cutoff.
First, $S(\rho)$ is computed from the eigenvalues of the $2\times 2$ matrix $\rho$ with entries
$\rho_{ij}=\langle \phi_i|\rho|\phi_j\rangle
$ in the L\"owdin-orthonormalized basis $\Phi=\Psi G^{-1/2}$ of
$\Span\{\ket{g_1},\ket{g_2}\}$, where $G_{ij}=\braket{g_i}{g_j}$.
Second, $S(\tau(\rho))$ depends only on the covariance matrix of $\rho$, which we obtain in closed form from the
Gaussian branch data and cross moments (\hyperref[appendixC]{Appendix~C}). In particular, for $p=0$ one recovers the pure-superposition limit
$S(\rho_0)=0$ and hence $\delta_{\rm nG}(\rho_0)=S(\tau(\rho_0))$.

\begin{figure}[h]
    \centering
    \includegraphics[width=\linewidth]{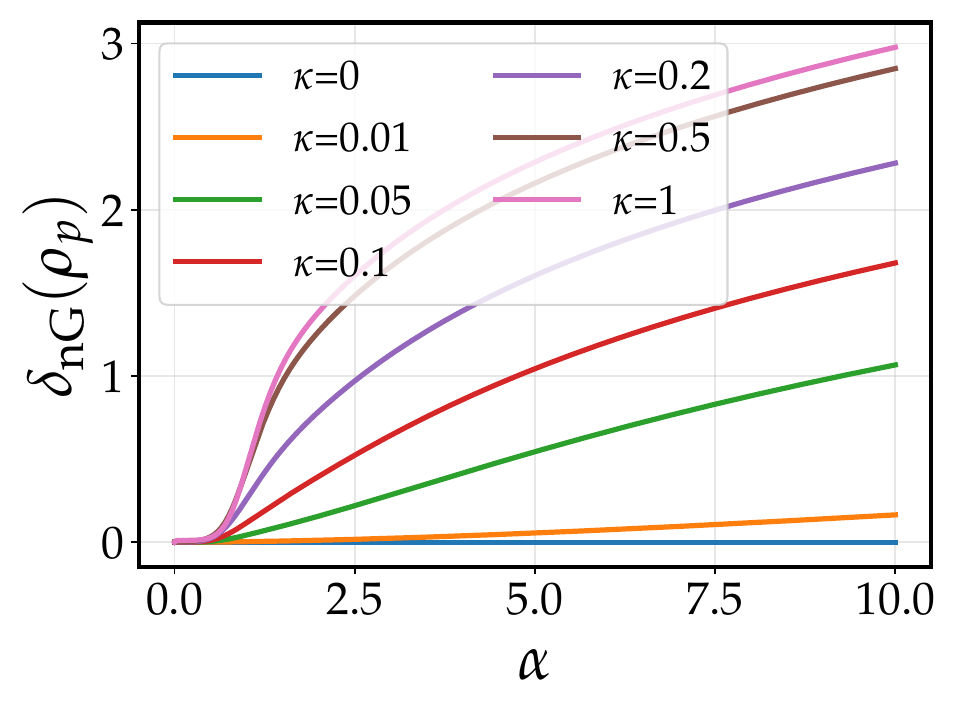}
 \caption{
Relative entropy non-Gaussianity
$\delta_{\rm nG}(\rho_p)=S(\tau(\rho_p))-S(\rho_p)$ for phase-randomized
two-branch single-mode states. The two Gaussian branches are symmetrically
displaced vacuum states,
$\ket{g_{1,2}}=D(\pm\alpha)\ket{0}$, while the superposition itself can be
uneven:
$\ket{\psi_\pm}\propto \ket{g_1}\pm\kappa\ket{g_2}$.
We fix the phase-mixture probability $p=0.1$ in Eq.~\eqref{eq:phase_randomization_model}
and evaluate $\delta_{\rm nG}$ as a function of the displacement $\alpha$ for several
branch-imbalance parameters
$\kappa\in\{0,0.01, 0.05,0.1,0.2,0.5,1\}$.
Here $\kappa=0$ gives a single Gaussian branch, while $\kappa=1$ gives the balanced
cat. The entropy $S(\rho_p)$ is obtained from the exact $2\times2$ matrix on
$\Span\{\ket{g_1},\ket{g_2}\}$ determined by Gaussian overlaps
(\hyperref[appendixB]{Appendix~B}), while $S(\tau(\rho_p))$ is obtained from the
covariance matrix computed analytically from Gaussian cross moments
(\hyperref[appendixC]{Appendix~C}).
}
    \label{fig:delta_nG}
\end{figure}

Figure~\ref{fig:delta_nG} shows $\delta_{\rm nG}(\rho_p)$ for phase-randomized uneven cat
superpositions at fixed $p=0.1$. This family probes a numerically delicate regime:
Changing $\kappa$ interpolates between a single coherent state and an even cat state, while increasing $\alpha$ separates the
branches in phase space. The curves are obtained exactly, whereas Fock-space truncations can give cutoff-dependent estimates in large-displacement regimes. A complementary operational analysis of the non-Gaussianity of Gaussian
mixtures, based on robustness measures, is given in Ref.~\cite{Turner2025}.

\paragraph{Negativity from the overlaps.}
\label{par:two_branch_entanglement}

We now consider two bosonic subsystems \(A\) and \(B\), each of which may contain several
modes. Let \(\ket{g_1^A},\ket{g_2^A}\) be two pure Gaussian states on \(A\), and
\(\ket{g_1^B},\ket{g_2^B}\) two pure Gaussian states on \(B\). We define the product
branches
\begin{equation}
\ket{\Gamma_1}:=\ket{g_1^A}\otimes\ket{g_1^B},
\qquad
\ket{\Gamma_2}:=\ket{g_2^A}\otimes\ket{g_2^B}.
\end{equation}
Such two-branch structures arise whenever a bosonic system becomes correlated with a
discrete degree of freedom, such as an ancilla qubit, detector outcome, or measurement
record, and also when nonlinear dynamics produces well-separated semiclassical
configurations whose interference is only partially preserved. Examples include
relativistic detector-field scenarios such as entanglement harvesting with
Unruh--DeWitt detectors
\cite{Reznik2005,fuentes2005alice,PozasKerstjens2015,foo2021thermality,BarbadoCastroRuizApadulaBrukner2020},
and nonlinear oscillator or photonic-network settings where Kerr or parametric
interactions generate cat-like branches and multimode superpositions
\cite{Inagaki2016,McMahon2016,grimm2020stabilization}.

We study the Bell-like Gaussian-branch superposition
\begin{equation}
\ket{\Psi_\varphi}
=
\frac{1}{\sqrt{\mathcal Z_\varphi}}
\Big(\ket{\Gamma_1}+e^{i\varphi}\ket{\Gamma_2}\Big),
\qquad
\mathcal Z_\varphi
=
2+2\,\Re\!\big(e^{i\varphi}ab\big),
\label{eq:bell}
\end{equation}
where the local branch overlaps are
\begin{equation}
a:=\braket{g_1^A}{g_2^A},
\qquad
b:=\braket{g_1^B}{g_2^B}.
\label{eq:local_overlaps_def}
\end{equation}
For two branches, local rephasings allow us to take \(a,b\ge0\), so that
\(\mathcal Z_\varphi=2+2ab\cos\varphi\).

The state \(\rho_\varphi=\ket{\Psi_\varphi}\!\bra{\Psi_\varphi}\) is supported on
\[
\Span\{\ket{g_1^A},\ket{g_2^A}\}
\otimes
\Span\{\ket{g_1^B},\ket{g_2^B}\},
\]
and therefore has an exact representation on an effective
\(\mathbb C^2\otimes\mathbb C^2\) logical space. In the local L\"owdin bases, the
partial transpose is the ordinary matrix partial transpose on the \(B\) logical index,
\begin{equation}
(\rho^{T_B})_{(i,\alpha),(j,\beta)}
=
\rho_{(i,\beta),(j,\alpha)},
\qquad
i,j,\alpha,\beta\in\{1,2\}.
\label{eq:PT_index_swap}
\end{equation}
Thus, for this finite branch-factorized manifold, the entanglement calculation is a finite-dimensional PPT calculation on the
encoded manifold; it does not require a phase-space partial transpose
\((x,p)\mapsto(x,-p)\). The finite representation also gives the Schmidt vectors: eigenvectors of the
effective reduced density matrices are mapped back to bosonic Schmidt modes by the
local L\"owdin bases.

\begin{figure}[h]
    \centering
    \includegraphics[width=\linewidth]{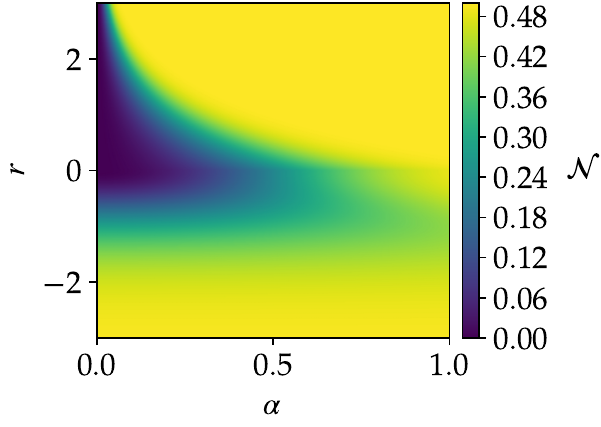}
\caption{
Negativity of a bosonic encoded Bell-like state
$\ket{\Psi_\varphi}\propto \ket{g_1}_A\ket{g_1}_B+e^{i\varphi}\ket{g_2}_A\ket{g_2}_B$,
evaluated in closed form from the local overlaps $a=\langle g_1|g_2\rangle_A$ and $b=\langle g_1|g_2\rangle_B$ as
$\mathcal N=\sqrt{(1-a^2)(1-b^2)}/\big(2(1+ab\cos\varphi)\big)$ ($a,b\ge0$).
The branches are displaced squeezed states (single mode per party), so that
$a=b$ follows directly from the pure-Gaussian fidelity in terms of first and second moments.
The plot shows \(\mathcal N\) as a function of displacement amplitude
\(\alpha\) and squeezing \(r\) for the in-phase case \(\varphi=0\);
for $r>0$ the two branches are squeezed along the same quadrature, while for
$r<0$ they are squeezed along conjugate quadratures.
}
    \label{fig:negativity}
\end{figure}

For the pure state~\eqref{eq:bell}, the negativity is
\begin{equation}
\mathcal N(\Psi_\varphi)
=
\frac{\sqrt{(1-a^2)(1-b^2)}}{2(1+ab\cos\varphi)} .
\label{eq:negativity}
\end{equation}
For this two-branch family, Eq.~\eqref{eq:negativity} recovers the general nonorthogonal-state entanglement relation~\cite{wang2002nonorthogonal}, now evaluated directly from arbitrary multimode Gaussian branch data. At fixed relative phase, the entanglement depends only on the two local overlaps. Thus, all information about
displacements, squeezing, phases, and the number of modes is compressed into
the local overlaps \(a\) and \(b\). Unlike a Bell state written in an
orthonormal discrete-variable basis, an equal-amplitude, Bell-like
superposition of two bosonic product branches is not generically maximally
entangled. For the in-phase superposition \(\varphi=0\), local
non-orthogonality suppresses entanglement, and the maximally entangled
two-qubit value \(\mathcal N=1/2\) is recovered only in the
orthogonal-branch limit \(a,b\to0\). For general phases, however, branch
interference can compensate local non-orthogonality. In particular, for the
antisymmetric superposition \(\varphi=\pi\) with equal local overlaps
\(a=b<1\), Eq.~\eqref{eq:negativity} gives
\(\mathcal N=1/2\) exactly, as for the entangled coherent state obtained by sending an odd cat
through a balanced beam splitter~\cite{vanEnkHirota2001}. Thus Eq.~\eqref{eq:negativity}
does not merely quantify the loss of entanglement due to non-orthogonality:
it also identifies when phase and overlap combine to produce a maximally
entangled logical qubit inside the infinite-dimensional bosonic Hilbert
space. Figure~\ref{fig:negativity} illustrates this dependence for displaced
squeezed branches.

Equivalently, since the state has Schmidt rank at most two, Eq.~\eqref{eq:negativity}
also determines the nonzero Schmidt spectrum. If \(\lambda_{1,2}\) are the two
nonzero eigenvalues of \(\rho_A=\Tr_B\ket{\Psi_\varphi}\!\bra{\Psi_\varphi}\), then
\begin{equation}\label{eq:schmidt}
\lambda_{1,2}
=
\frac12\left(1\pm\sqrt{1-4\,\mathcal N(\Psi_\varphi)^2}\right),
\end{equation}
with all other eigenvalues equal to zero in the full bosonic Hilbert space.
The same finite-support construction also treats dephased branch mixtures:
the explicit \(4\times4\) logical matrix and the closed form formula are given in \hyperref[app:dephased_bell_negativity]{Appendix D}.

Crucially, Eq.~\eqref{eq:negativity} provides an analytic entanglement evaluation
for highly excited multimode non-Gaussian bosonic states using only Gaussian overlaps and finite-dimensional linear
algebra, avoiding any Fock-space truncation.
The same overlap-based structure also provides a benchmark for many-body entanglement. Appendix E derives the exact entropy scaling of Gaussian-branch GHZ states, showing how finite branch overlap controls the crossover from nonorthogonal bosonic cats to ideal discrete-variable GHZ behavior, and exposing Fock-truncation errors in bright-displaced regimes.

\paragraph{Discussion.}
Gaussian superpositions have become a practical language for non-Gaussian bosonic physics. Cat states and related Gaussian-branch encodings are now routinely generated and stabilized in cavity and circuit QED, appear in optical and microwave bosonic error correction, and provide a natural description of branch-conditioned dynamics in detector--field and light--matter settings. Recent simulation methods have shown that such states are not merely formal ans\"atze: linear combinations of Gaussian functions make bosonic-qubit circuits and Gaussian-branch dynamics accessible in regimes where direct Fock-space methods are impractical~\cite{bourassa2021fast,marshall2023simulation,dias2024classical,hahn2025classical,solodovnikova2025fast,braccini2025superpositions}. In relativistic field-theoretic detector models, coherent Gaussian superpositions have similarly enabled the treatment of regimes that were previously inaccessible with standard methods~\cite{funai2025gaussian}. 

Recent developments further show that finite Gaussian-branch structures are
not only useful for few-mode encodings, but also arise as organizing principles
for many-body and resource-state constructions. Gaussian-augmented bosonic
matrix-product states combine Gaussian structure with finite bond dimension,
providing a variational tensor-network framework for bosonic many-body systems
without imposing a local Fock cutoff~\cite{tjoa2026gaussianaugmented}.
Complementarily, parent-Hamiltonian constructions for entangled bosonic cat
resources show how coherent-branch manifolds, together with GHZ-, cluster-, or
W-type correlations, can be selected directly by physical constraints on
oscillator modes~\cite{brauer2026symmetryresolved}. These developments
reinforce the motivation for the present work: once a bosonic state is
represented inside a finite nonorthogonal branch manifold, its quantum-
information content requires spectral diagnostics, not only expectation values.

The present work supplies the complementary spectral layer. Propagating or representing a Gaussian-branch state does not, by itself, give the eigenvalues, trace norms, entropies, or negativities of the resulting non-Gaussian infinite-dimensional operator. We showed that, whenever the state is supported on a finite family of pure Gaussian branches, these spectral quantities are instead obtained from an exact finite logical density matrix. The oscillator or field degrees of freedom remain infinite-dimensional; the finite object arises from the branch support, not from a photon-number cutoff.

This gives direct access to information-theoretic quantities that are otherwise difficult to evaluate beyond the Gaussian regime. For mixed two-branch states, the von Neumann entropy is obtained from the finite logical spectrum, while the Gaussian reference entropy entering relative entropy non-Gaussianity is computed from Gaussian cross moments. In this way, Gaussian-branch representations, previously used mainly for propagation, sampling, or expectation values, become tools for entropy-based diagnostics of mixed non-Gaussian bosonic states.

The same viewpoint yields a compact entanglement characterization for a technologically relevant family of bosonic states. For arbitrary multimode two-branch product Gaussian superpositions, the encoded bipartite negativity depends only on the local Gaussian overlaps. Thus, the continuous-variable details of the branches (displacements, squeezing, phases, mode number, and energy) enter the entanglement only through branch distinguishability. This provides a cutoff-free benchmark for cat-like entanglement in quantum optics, superconducting circuits, and hybrid mechanical or detector-field platforms, and generalizes the familiar entangled-coherent-state picture to arbitrary pure Gaussian branches. As a many-body illustration, Appendix E shows that the finite logical spectrum also gives analytic entanglement-entropy scaling for Gaussian-branch GHZ states, connecting cat-like oscillator resources with standard discrete-variable entanglement diagnostics in a cutoff-free way. Together, these examples show that the finite logical spectrum gives access
not only to single-state entropy diagnostics, but also to bipartite and
many-body entanglement structure in nonorthogonal bosonic cat manifolds.

The range of potential applications is broad. In superconducting circuits, stabilized cat and Kerr-cat qubits provide a natural setting where Gaussian-branch logical manifolds are engineered directly~\cite{Vlastakis2013Cat,leghtas2015confining,ofek2016extending,grimm2020stabilization}. In strong-field and attosecond quantum optics, bright squeezed or nonclassical driving fields have recently been shown to generate Gaussian entanglement and non-Gaussian optical features in high-harmonic and above-threshold-ionization processes~\cite{rivera2025microscopic}. In levitated and nanomechanical systems, spatial or phononic cat-like branches arise naturally from conditional forces, electromechanical coupling, or postselection~\cite{romero2011large,riedinger2018remote,ockeloen2018stabilized,martinetz2020quantum}. In all these cases, the relevant physical state may be described compactly by Gaussian data while its quantum-information content is spectral and non-Gaussian.

Several extensions follow naturally. Larger branch manifolds allow one to treat qudit encodings, symmetry-generated Gaussian orbits, and finite grid-state approximations. The same finite-support construction can be used for distinguishability and hypothesis-testing quantities when the states being compared share the selected branch manifold, and for further encoded-state entanglement diagnostics. It also provides a natural reduced objective for variational design: optimization can be performed on the finite spectral object while the physical implementation remains specified by Gaussian branch parameters~\cite{stornati2024variational}.

The construction also has clear boundaries. It is exact only while the state remains in the chosen finite branch manifold; generic dynamics and noise need not preserve this support unless the branch set is enlarged or the channel has a compatible finite-branch action. Similarly, the finite PPT calculation applies when the branch manifold factorizes across the chosen bipartition. For arbitrary spatial or mode cuts, partial transposition and tracing can lead to finite sums of Gaussian transition operators rather than finite-rank branch states. Extending the present spectral reduction to such Gaussian-operator sums, for example through R\'enyi and partial-transpose moments, is a natural direction for open dynamics and field-theoretic applications.

Overall, the finite Gaussian-branch density matrix identifies a broad tractable sector of non-Gaussian bosonic Hilbert space. It turns physically natural Gaussian-superposition descriptions into exact finite spectral objects, providing a cutoff-free route to quantify entropy, non-Gaussianity, and entanglement in bosonic encodings and related oscillator or field platforms.

\paragraph{Acknowledgements.} F.C. acknowledges funding by the European Union (EQC, 101149233). The authors thank Beatriz Polo, Mireia Egidio and Iris Julian for fruitful feedback.

\bibliographystyle{apsrev4-2}
\bibliography{biblio}

\appendix

\onecolumngrid

\section{Appendix A: Orthogonalization}
\label{app:finite_manifold}

Let $\mathcal H$ be a Hilbert space and let
$\mathcal B=\{\ket{\psi_k}\}_{k=0}^{D-1}\subset\mathcal H$
be a linearly independent family of normalized (generally non-orthogonal) vectors.
Define the $D\times D$ Gram matrix
\begin{equation}
G_{ij}:=\braket{\psi_i}{\psi_j},\qquad G=G^\dagger,\qquad G\succ 0.
\label{eq:app_gram}
\end{equation}
Collect the kets into the (formal) column operator
\begin{equation}
\Psi:=\big[\,\ket{\psi_0}\ \cdots\ \ket{\psi_{D-1}}\,\big],
\qquad
\text{so that}\quad \Psi^\dagger\Psi=G .
\label{eq:app_Psi_def}
\end{equation}

A convenient orthonormal basis of $\Span(\mathcal B)$ is obtained by symmetric orthogonalization:
\begin{equation}
\Phi:=\Psi\,G^{-1/2}
=
\big[\,\ket{\phi_0}\ \cdots\ \ket{\phi_{D-1}}\,\big].
\label{eq:app_Phi_def}
\end{equation}
Then
\begin{equation}
\Phi^\dagger\Phi
=
G^{-1/2}\Psi^\dagger\Psi\,G^{-1/2}
=
G^{-1/2}GG^{-1/2}
=
I_D,
\label{eq:app_isometry}
\end{equation}
so the columns of $\Phi$ form an orthonormal basis of $\Span(\mathcal B)$.
Moreover,
\begin{equation}
\Psi=\Phi\,G^{1/2}.
\label{eq:app_Psi_factor}
\end{equation}
Any other orthonormal basis $\Phi'=\Phi U$ with $U\in U(D)$ yields unitarily equivalent finite matrices
and hence the same eigenvalues; all spectral quantities are basis independent.

After introducing these bases, we now express arbitrary states and operators in terms of them. Consider now an arbitrary density operator supported on $\Span(\mathcal B)$, written as
\begin{equation}\label{eq:app_rho_C}
\rho
=
\sum_{i,j=0}^{D-1} C_{ij}\ketbra{\psi_i}{\psi_j}
=
\Psi\,\mathbf C\,\Psi^\dagger .
\end{equation}
Where $\mathbf C$ is a $D\times D$ matrix with elements $C_{ij}$. The conditions that $\rho$ is Hermitian, positive, and normalized imply that
\begin{equation}
\mathbf C=\mathbf C^\dagger,\qquad
G^{1/2}\mathbf C G^{1/2}\succeq 0,\qquad
\Tr(\mathbf C G)=1.
\label{eq:app_C_conditions}
\end{equation}
Using Eq.~\eqref{eq:app_Psi_factor}, the same state can be written in the orthonormal basis $\Phi$ as
\begin{equation}
\rho
=
\Phi\,
\left(G^{1/2}\mathbf C G^{1/2}\right)
\Phi^\dagger .
\label{eq:app_rho_matrix}
\end{equation}
The map $\Phi:\mathbb C^D\to\mathcal H$ is an isometry onto $\Span(\mathcal B)$ by
Eq.~\eqref{eq:app_isometry}; hence this finite matrix has the same nonzero spectrum as the original
operator on $\mathcal H$. Therefore, all quantities depending only on the nonzero spectrum
reduce to standard $D\times D$ linear algebra.

Any pure state in the manifold can be written as
\begin{equation}
\ket*{\psi^{(\mathbf c)}}=\sum_{k=0}^{D-1} c_k\ket{\psi_k}=\Psi\,\mathbf c,
\qquad \mathbf c\in\mathbb C^D,
\label{eq:app_pure_state}
\end{equation}
with norm
\begin{equation}
\braket*{\psi^{(\mathbf c)}}{\psi^{(\mathbf c)}}=\mathbf c^\dagger G \mathbf c.
\label{eq:app_norm}
\end{equation}
In the orthonormal basis $\Phi$ it is represented by the coefficient vector $\bf \tilde c$:
\begin{equation}
\ket*{\psi^{(\mathbf c)}}=\Phi\,\tilde{\mathbf c},
\qquad
\tilde{\mathbf c}:=G^{1/2}\mathbf c.
\label{eq:app_coeff_map}
\end{equation}
The state is normalized iff $\mathbf c^\dagger G\mathbf c=1$, and in that case the corresponding
coefficient matrix in Eq.~\eqref{eq:app_rho_C} is
\begin{equation}
\mathbf C=\mathbf c\,\mathbf c^\dagger.
\label{eq:app_pure_C}
\end{equation}

\noindent\emph{Characteristic function:}
\label{par:gaussian_phasespace}
In this work $|\psi_k\rangle$ are (pure) $n$-mode Gaussian states
 $|g_k\rangle$, specified by
first moments $d_k\in\mathbb R^{2n}$ and covariance matrices $V_k\in\mathbb R^{2n\times 2n}$,
with vacuum covariance $V_{\rm vac}=\tfrac12 I$. Let
\begin{equation}
R=(\hat x_1,\ldots,\hat x_n,\hat p_1,\ldots,\hat p_n)^{\mathsf T},\qquad [R_a,R_b]=i\Omega_{ab},
\end{equation}
where $\Omega=\begin{psmallmatrix}0&I_n\\-I_n&0\end{psmallmatrix}$ is the symplectic form in this grouped ordering. We use the Weyl displacement operator
\begin{equation}
D(\xi):=\exp\!\big(i\,\xi^{\mathsf T}\Omega R\big),\qquad \xi\in\mathbb R^{2n},
\label{eq:weyl_displacement}
\end{equation}
and the characteristic function of an operator $A$ defined as:
\begin{equation}
\chi_A(\xi):=\Tr\!\big[A\,D(\xi)\big].
\label{eq:charfun_def}
\end{equation}
For a pure Gaussian state $(d,V)$ one has a Gaussian characteristic function:
\begin{equation}
\chi_G(\xi)=\exp\!\left(-\frac12\,\xi^{\mathsf T}\Omega V\Omega^{\mathsf T}\xi
+i\,\xi^{\mathsf T}\Omega d\right).
\label{eq:gaussian_charfun}
\end{equation}

Let us introduce the cross characteristic function for  non-orthogonal Gaussian basis as:
\begin{equation}
\chi_{ij}(\xi):=\langle g_i|D(\xi)|g_j\rangle,
\qquad
\chi_{ij}(0)=\langle g_i|g_j\rangle\equiv G_{ij}.
\label{eq:cross_charfun_def}
\end{equation}
Hence the Gram matrix is obtained as $G_{ij}=\chi_{ij}(0)$.
Now, we can we define the matrix $\bm \chi(\xi)$ with elements $\chi_{ij}(\xi)$, and the characteristic function of a state $\ket{\psi^{(\bf c)}}$ reads:
\begin{equation}
\chi_{\psi}(\xi):=\langle\psi^{(\bf c)}|D(\xi)|\psi^{(\bf c)}\rangle
= \mathbf{c^{\dagger} \bm\chi(\xi) \ \bf c}
\label{eq:charfun_superposition}
\end{equation}
For an arbitrary density matrix $\rho$ as in Eq.~\eqref{eq:app_rho_C} the characteristic function reads:
\begin{equation}
\chi_{\rho}(\xi)
= {\rm Tr}(\mathbf{C\, {\bm\chi}(\xi))}
\end{equation}
This construction extends to arbitrary operators. That is for a given operator $A$ we can define the  matrix $\mathbf A$ with elements $A_{ij}=\bra{g_i}A\ket{g_j}$ and the mean value for a states with support in this subspace can be expressed as:
\begin{equation}
{\rm Tr}(\rho A)={\rm Tr}(\mathbf C\mathbf A)
\end{equation}

\noindent\emph{Consistency and physicality conditions:}
\label{par:physicality}
The diagonal characteristic functions 
$\chi_{jj}(\xi)=\Tr\!\big[|g_j\rangle\langle g_j|\,D(\xi)\big]$
describe physical Gaussian states and therefore satisfy the usual uncertainty relation
\begin{equation}
V_j+\frac{i}{2}\Omega\succeq 0,
\qquad
\det V_j=2^{-2n}\ \text{ for pure states}.
\label{eq:uncertainty_diagonal}
\end{equation}
By contrast, for $i\neq j$ the operators $\sigma_{ij}:=|g_i\rangle\langle g_j|$ are neither Hermitian
nor positive, hence $\Tr\!\big[\sigma_{ij}D(\xi)\big]=\chi_{ji}(\xi)$ is not the characteristic function
of a quantum state and is not required to satisfy Eq.~\eqref{eq:uncertainty_diagonal} by itself.
Instead, off-diagonal blocks obey \emph{consistency} constraints inherited from their definition.

First, Hermitian conjugation implies
\begin{equation}
\sigma_{ji}=\sigma_{ij}^\dagger,
\qquad
D(\xi)^\dagger=D(-\xi),
\end{equation}
and therefore the cross characteristic functions satisfy
\begin{equation}
 {\ \chi_{ji}(\xi)=\chi_{ij}(-\xi)^* \ },
\label{eq:chi_hermiticity}
\end{equation}
in particular $G_{ji}=G_{ij}^*$ and $G_{ii}=1$.
Second, since $D(\xi)$ is unitary and $\|\sigma_{ij}\|_1=\||g_i\rangle\langle g_j|\|_1=1$, one has the
uniform bound
\begin{equation}
 {\ |\chi_{ij}(\xi)|\le 1 \quad \forall\,\xi\in\mathbb R^{2n}. \ }
\label{eq:chi_bound}
\end{equation}
Third, the Gram matrix $G=[G_{ij}]$ is positive semidefinite for any family of kets; in our setting
we assume $G\succ 0$ (linear independence) so that the L\"owdin map $G^{-1/2}$ is well defined.


\section{Appendix B: Information-theoretic quantities.}
\label{appendixB}

All single-party information measures considered below reduce to finite-dimensional linear algebra.
Indeed, any admissible state supported on $\Span(\mathcal B)$ is represented in the orthonormal
L\"owdin basis $\{|\phi_i\rangle\}$ by the $D\times D$ density matrix $\rho$ in
Eq.~\eqref{eq:app_rho_matrix}, defined in the previous Appendix. Denoting by $\{\lambda_a\}_{a=1}^D$ the eigenvalues of $\rho$,
the von Neumann and R\'enyi entropies are
\begin{align}
S(\rho) &=-\Tr(\rho\log\rho) = -\sum_{a=1}^D \lambda_a\log\lambda_a,
\label{eq:S_from_spectrum}\\
S_{\alpha}(\rho) &=\frac{1}{1-{\alpha}}\log\Tr(\rho^{\alpha})
=\frac{1}{1-{\alpha}}\log\!\left(\sum_{a=1}^D \lambda_a^{\alpha}\right),
\; {\alpha}>0,\ {\alpha}\neq 1.
\label{eq:Renyi_from_spectrum}
\end{align}

The eigenvalues $\{\lambda_a\}$ may be obtained by diagonalizing $\rho$ in eq. \ref{eq:app_rho_matrix}, or,
equivalently, by solving the generalized eigenvalue problem
\begin{equation}
G\mathbf C G\,u=\lambda G\,u,
\label{eq:gen_eig_XG}
\end{equation}
which avoids constructing an explicit orthonormal basis in $\mathcal H$.

We quantify non-Gaussianity by the relative entropy distance to the reference Gaussian state \cite{Genoni2008PRA_relativeEntropyNG,GenoniParis2010}
$\tau(\rho)$ having the same first and second moments $(d,V)$ as $\rho$,
\begin{equation}
\delta_{\mathrm{nG}}(\rho):=S\!\big(\rho\|\tau(\rho)\big).
\label{eq:delta_nG_def}
\end{equation}
Since $\tau(\rho)$ is the maximum-entropy state compatible with $(d,V)$, one has the identity
\begin{equation}
\delta_{\mathrm{nG}}(\rho)=S(\tau(\rho)) - S(\rho),
\label{eq:delta_nG_Sdiff}
\end{equation}
where $S(\rho)$ is obtained from $\{\lambda_a\}$ via Eq.~\eqref{eq:S_from_spectrum} and
$S(\tau(\rho))$ depends only on the symplectic eigenvalues $\{\nu_k\}_{k=1}^n$ of $V$:
\begin{equation}
S(\tau(\rho))=\sum_{k=1}^n
\left[
\left(\nu_k+\frac12\right)\log\!\left(\nu_k+\frac12\right)
-
\left(\nu_k-\frac12\right)\log\!\left(\nu_k-\frac12\right)
\right].
\label{eq:gaussian_entropy_symplectic}
\end{equation}

\noindent\emph{Entropy of a noisy non-Gaussian state:}
\label{app:entropy}
Let $\ket{g_1},\ket{g_2}$ be two pure Gaussian branches with overlap
\begin{equation}
g:=\braket{g_1}{g_2},\qquad 0\le g<1,
\end{equation}
where we fixed the relative phase of $\ket{g_2}$ so that $g$ is real and nonnegative.
In the non-orthogonal branch basis $\{\ket{g_1},\ket{g_2}\}$ the Gram matrix is
\begin{equation}
G=\begin{pmatrix}1 & g\\ g & 1\end{pmatrix}.
\label{eq:Gram_two_branch_app}
\end{equation}
Consider the real-amplitude superposition with amplitude ratio $\kappa\ge 0$,
\begin{equation}
\ket{\Psi_{+}}=\frac{1}{\sqrt{\mathcal Z_{+}}}\Big(\ket{g_1}+\kappa\,\ket{g_2}\Big),
\qquad
\mathcal Z_{+}=1+\kappa^2+2\kappa g,
\label{eq:Psi_plus_def_app}
\end{equation}
and its phase-flipped partner (relative phase $\pi$),
\begin{equation}
\ket{\Psi_{-}}=\frac{1}{\sqrt{\mathcal Z_{-}}}\Big(\ket{g_1}-\kappa\,\ket{g_2}\Big),
\qquad
\mathcal Z_{-}=1+\kappa^2-2\kappa g.
\label{eq:Psi_minus_def_app}
\end{equation}
Picking one of the two possible phases with probability $p$ produces the mixed state
\begin{equation}
\rho:=(1-p)\ket{\Psi_{+}}\!\bra{\Psi_{+}}+p\,\ket{\Psi_{-}}\!\bra{\Psi_{-}},
\qquad 0\le p\le 1.
\label{eq:rho_phase_random_app}
\end{equation}
This phenomenological model describes preparation uncertainty between the two relative phases while keeping the support inside $\Span\{\ket{g_1},\ket{g_2}\}$. It is not, for general nonorthogonal branches, the action of a deterministic branch-selective phase-flip unitary.

Let $\Phi=\Psi\,G^{-1/2}$ be the L\"owdin orthonormalization of the branch family,
with $\Psi=[\ket{g_1}\ \ket{g_2}]$. We represent the state on its support by its $2\times 2$ matrix in the orthonormal basis $\Phi$. The effective $2\times 2$ density matrix is
\begin{equation}
\rho=\Phi\left((1-p)\,\tilde c_{+}\tilde c_{+}^\dagger+p\,\tilde c_{-}\tilde c_{-}^\dagger\right)\Phi^\dagger,
\label{eq:rho_phi_phase_random_app}
\end{equation}
where the \emph{orthonormal-basis coefficient vectors} are obtained directly from the Gram matrix,
\begin{equation}
\tilde c_{\pm}=\frac{G^{1/2}c_{\pm}}{\sqrt{c_\pm^\dagger G c_\pm}},
\qquad
c_{+}=\binom{1}{\kappa},\quad c_{-}=\binom{1}{-\kappa}.
\label{eq:tilde_c_pm_app}
\end{equation}
For the symmetric Gram matrix~\eqref{eq:Gram_two_branch_app}, a convenient closed form is
\begin{equation}
G^{1/2}
=
\frac12
\begin{pmatrix}
a+b & a-b\\
a-b & a+b
\end{pmatrix},
\qquad
a:=\sqrt{1+g},\quad b:=\sqrt{1-g},
\label{eq:Ghalf_closed_app}
\end{equation}
which follows from diagonalizing $G$ in the symmetric/antisymmetric eigenbasis.

Since $\rho$ is $2\times 2$, its eigenvalues are
\begin{equation}
\lambda_{\pm}
=
\frac12\left(1\pm\sqrt{1-4\det\rho}\right),
\label{eq:eigs_2x2_app}
\end{equation}
and the von Neumann entropy is $S(\rho)=-\sum_{\pm}\lambda_\pm\log\lambda_\pm$.

A direct evaluation from~\eqref{eq:rho_phi_phase_random_app}--\eqref{eq:Ghalf_closed_app} gives the determinant
in \emph{closed form}:
\begin{equation}
{\det\rho
=
\frac{4\,p(1-p)\,\kappa^2(1-g^2)}{(1+\kappa^2)^2-(2\kappa g)^2}.
}
\label{eq:det_rho_phi_app}
\end{equation}
Equations~\eqref{eq:eigs_2x2_app}--\eqref{eq:det_rho_phi_app} therefore provide an analytic expression for
$S(\rho)$ \emph{without any Fock-space truncation}.

For the balanced superposition $\kappa=1$, one has
\begin{equation}
\det\rho=p(1-p),
\end{equation}
so the spectrum $\{\lambda_\pm\}$ depends \emph{only} on $p$ and becomes independent of the branch overlap $g$.

\section{Appendix C: Moments mapping}\label{appendixC}
Throughout this appendix, branches are pure Gaussians, so $\det V=2^{-2n}$ and $V_{xx}\succ0$. We use the same grouped quadrature ordering as in Appendix A: $R=(\hat x_1,\ldots,\hat x_n,\hat p_1,\ldots,\hat p_n)^{\mathsf T}$, with $[\hat x_j,\hat p_k]=i\delta_{jk}$.
To evaluate $\chi_{ij}(\xi)$ in closed form we fix a global phase convention for Gaussian kets. Care is required when fixing a consistent phase convention for pure-Gaussian overlaps~\cite{yao2024riemannian,dias2024classical,quesada2025s,sun2026representation,martinez2026calculating}. 
A convenient choice in this case is the canonical position-space gauge in which the wavefunction prefactor is
real and positive. Writing $\xi=(u,v)^{\mathsf T}$ with $u,v\in\mathbb R^n$, and partitioning the
covariance matrix as
\begin{equation}
V=
\begin{pmatrix}
V_{xx} & V_{xp}\\
V_{px} & V_{pp}
\end{pmatrix},
\qquad V_{px}=V_{xp}^{\mathsf T},
\end{equation}
the corresponding (pure) Gaussian wavefunction in the canonical position-space gauge can be written as
\begin{equation}
\psi(x)\equiv\langle x|g\rangle
=\mathcal N\exp\!\left(-\frac12 x^{\mathsf T}A\,x+\beta^{\mathsf T}x+\gamma\right),
\qquad x\in\mathbb R^n,
\label{eq:wf_A_beta_gamma}
\end{equation}
with parameters determined by moments $(d,V)$ as
\begin{equation}
\mathcal N=(\det(2\pi V_{xx}))^{-1/4},\qquad
A=V_{xx}^{-1}\!\left(\frac12 I - i V_{xp}\right),\qquad
\beta=A x_0 + i p_0,\qquad
\gamma=-\frac12 x_0^{\mathsf T}A x_0-\frac{i}{2}p_0^{\mathsf T}x_0,
\label{eq:A_beta_gamma_from_moments}
\end{equation}
where $d=(x_0,p_0)^{\mathsf T}$.

In the position representation the Weyl operator acts as
\begin{equation}
\big[D(u,v)\psi\big](x)=\exp\!\big(-i\,v^{\mathsf T}(x+\tfrac12 u)\big)\,\psi(x+u),
\label{eq:D_action_xrep_clean}
\end{equation}
which maps $(A,\beta,\gamma)\mapsto(A,\beta(u,v),\gamma(u,v))$ with
\begin{align}
\beta(u,v)&:=\beta - A u - i v, \label{eq:beta_shift}\\
\gamma(u,v)
&:=
\gamma+\beta^{\mathsf T}u
-\frac12u^{\mathsf T}Au
-\frac{i}{2}v^{\mathsf T}u .\label{eq:gamma_shift}
\end{align}
For $(A_i,\beta_i,\gamma_i,\mathcal N_i)$ and $(A_j,\beta_j,\gamma_j,\mathcal N_j)$ obtained from
$(d_i,V_i)$ and $(d_j,V_j)$ via Eq.~\eqref{eq:A_beta_gamma_from_moments}, Gaussian integration yields
\begin{equation}
 {
\ \chi_{ij}(u,v)
=
\mathcal N_i^*\mathcal N_j\,
(2\pi)^{n/2}\exp\!\left[-\tfrac12\operatorname{Tr}\operatorname{Log}(A_i^*+A_j)\right]\;
\exp\!\left[
\gamma_i^*+\gamma_j(u,v)+\frac12\big(\beta_i^*+\beta_j(u,v)\big)^{\mathsf T}
(A_i^*+A_j)^{-1}\big(\beta_i^*+\beta_j(u,v)\big)
\right],
}
\label{eq:chi_ij_closed_form_clean}
\end{equation}
where $\operatorname{Log}$ denotes the principal matrix logarithm, ensuring the analytic Gaussian-integral phase. Setting $(u,v)=(0,0)$ recovers the overlap
$G_{ij}=\chi_{ij}(0)$, whose modulus satisfies the standard fidelity identity
\begin{equation}
|\langle g_i|g_j\rangle|^2
=
\frac{1}{\sqrt{\det(V_i+V_j)}}
\exp\!\left[
-\frac12
(d_i-d_j)^{\mathsf T}(V_i+V_j)^{-1}(d_i-d_j)
\right].
\label{eq:gaussian_overlap_magnitude}
\end{equation}

\paragraph{Cross moments and covariance for Gaussian superpositions}
\label{app:cross_moments_cov}

This appendix collects explicit formulas for first and second moments of \emph{coherent superpositions}
of Gaussian branches, and their reduction to closed form in the two-branch single-mode case.

Let $\ket{g_1},\ket{g_2}$ be two pure Gaussian branches and
\begin{equation}
\ket{\Psi}=\frac{1}{\sqrt{\mathcal Z}}\big(c_1\ket{g_1}+c_2\ket{g_2}\big),\qquad
\mathcal Z=\|c_1\ket{g_1}+c_2\ket{g_2}\|^2.
\end{equation}
Define the overlap and cross moments
\begin{equation}
g_{12}=g:=\braket{g_1}{g_2}\in  \mathbb{C},\;\; |g|\leq 1,\qquad
r_{12}:=\bra{g_1}R\ket{g_2},\qquad
M_{12}:=\frac12\bra{g_1}\{R,R^{\mathsf T}\}\ket{g_2},
\end{equation}
with $r_{21}=r_{12}^*$ and $M_{21}=M_{12}^*$.
The normalization is
\begin{equation}
 {
\ \mathcal Z
=
|c_1|^2+|c_2|^2+c_1^*c_2\,g+c_2^*c_1\,g^* \ }.
\label{eq:Z_two_branch_app}
\end{equation}
Writing $d_k=\bra{g_k}R\ket{g_k}$ and $M_k=\frac12\bra{g_k}\{R,R^{\mathsf T}\}\ket{g_k}=V_k+d_k d_k^{\mathsf T}$,
the coherent-superposition moments are
\begin{align}
{
\ d
=
\frac{|c_1|^2 d_1+|c_2|^2 d_2+c_1^*c_2\,r_{12}+c_2^*c_1\,r_{12}^*}{\mathcal Z}
\ },
\label{eq:d_two_branch_app}\\
 {
\ M
=
\frac{|c_1|^2 M_1+|c_2|^2 M_2+c_1^*c_2\,M_{12}+c_2^*c_1\,M_{12}^*}{\mathcal Z}
\ },
\qquad
 {\ V=M-d\,d^{\mathsf T}\ }.
\label{eq:V_two_branch_app}
\end{align}
These formulas are fully explicit once $(g,r_{12},M_{12})$ are known.

In the canonical position-space gauge (real positive prefactor), each pure Gaussian branch is
\begin{equation}
\psi_k(x)=\braket{x}{g_k}
=\mathcal N_k\exp\!\left(-\frac12 A_k x^2+\beta_k x+\gamma_k\right),
\qquad k\in\{1,2\},\qquad \Re(A_k)>0,
\label{eq:wf_single_mode_app_rewrite}
\end{equation}
with parameters $(A_k,\beta_k,\gamma_k,\mathcal N_k)$ determined by the branch moments $(d_k,V_k)$ as in
Eq.~\eqref{eq:A_beta_gamma_from_moments}.

Define the scalar combinations
\begin{equation}
S:=A_1^*+A_2,\qquad B:=\beta_1^*+\beta_2,\qquad
\mu:=\frac{B}{S},\qquad \sigma:=\frac{1}{S}.
\label{eq:S_B_mu_sigma_app}
\end{equation}

The overlap is the Gaussian integral
\begin{equation}
{
\ g_{12}:=\braket{g_1}{g_2}
=
\mathcal N_1^*\mathcal N_2\,
\sqrt{\frac{2\pi}{S}}\;
\exp\!\left(\gamma_1^*+\gamma_2+\frac{B^2}{2S}\right)
\ }.
\label{eq:g12_closed_app}
\end{equation}

For $R=(\hat x,\hat p)^{\mathsf T}$ one finds
\begin{equation}
{
\ \bra{g_1}\hat x\ket{g_2}=g_{12}\,\mu,\qquad
\bra{g_1}\hat p\ket{g_2}=g_{12}\,(iA_2\mu-i\beta_2)\ }.
\label{eq:r12_closed_app}
\end{equation}

The required second moments are
\begin{align}
 {
\ \bra{g_1}\hat x^2\ket{g_2}=g_{12}\,(\sigma+\mu^2)\ },
\label{eq:x2_cross_app}\\
 {
\ \frac12\bra{g_1}\{\hat x,\hat p\}\ket{g_2}
=
g_{12}\left(iA_2(\sigma+\mu^2)-i\beta_2\mu-\frac{i}{2}\right)\ },
\label{eq:xp_cross_app}\\
 {
\ \bra{g_1}\hat p^2\ket{g_2}
=
g_{12}\left(A_2-A_2^2(\sigma+\mu^2)+2A_2\beta_2\mu-\beta_2^2\right)\ }.
\label{eq:p2_cross_app}
\end{align}
Therefore the cross symmetrized moment matrix $M_{12}=\tfrac12\bra{g_1}\{R,R^{\mathsf T}\}\ket{g_2}$ is
\begin{equation}
 {
\ M_{12}=
\begin{pmatrix}
\bra{g_1}\hat x^2\ket{g_2} & \tfrac12\bra{g_1}\{\hat x,\hat p\}\ket{g_2}\\[2pt]
\tfrac12\bra{g_1}\{\hat x,\hat p\}\ket{g_2} & \bra{g_1}\hat p^2\ket{g_2}
\end{pmatrix}
\ }.
\label{eq:M12_closed_app}
\end{equation}
Substituting \eqref{eq:g12_closed_app}--\eqref{eq:M12_closed_app} into
\eqref{eq:d_two_branch_app}--\eqref{eq:V_two_branch_app} yields fully explicit
first and second moments of an arbitrary two-branch single-mode Gaussian superposition.

When the branches become nearly linearly dependent, the Gram matrix becomes ill-conditioned and any
orthogonalization can be numerically unstable. In practice one may monitor the smallest eigenvalue of $G$
and, if needed, reduce the manifold size or use an SVD-based pseudoinverse with a controlled threshold.

\section{Appendix D: Bipartite entanglement}
\label{app:dephased_bell_negativity}

Let
\begin{equation}
a:=\braket{g_1^A}{g_2^A},\qquad
b:=\braket{g_1^B}{g_2^B},
\end{equation}
where the phases of the second branches have been fixed so that \(a,b\in[0,1]\).
Define
\begin{equation}
u:=\sqrt{\frac{1+a}{2}},\qquad
v:=\sqrt{\frac{1-a}{2}},\qquad
w:=\sqrt{\frac{1+b}{2}},\qquad
z:=\sqrt{\frac{1-b}{2}}.
\end{equation}
In the orthonormal product basis
\[
\{\ket{00},\ket{01},\ket{10},\ket{11}\}
=
\{\ket{0_A0_B},\ket{0_A1_B},\ket{1_A0_B},\ket{1_A1_B}\},
\]
the encoded Bell-like state
\begin{equation}
\ket{\Psi_\varphi}
=
\frac{\ket{\Gamma_1}+e^{i\varphi}\ket{\Gamma_2}}
{\sqrt{\mathcal Z_\varphi}},
\qquad
\mathcal Z_\varphi=2+2ab\cos\varphi,
\end{equation}
has density matrix \(\rho_\varphi=\ket{\Psi_\varphi}\bra{\Psi_\varphi}\) given by
\begin{equation}
\rho_\varphi
=
\frac{1}{\mathcal Z_\varphi}
\begin{pmatrix}
u^2w^2 C_+ & i u^2wz S & i uvw^2 S & uvwz C_+\\
-i u^2wz S & u^2z^2 C_- & uvwz C_- & -i uvz^2 S\\
-i uvw^2 S & uvwz C_- & v^2w^2 C_- & -i v^2wz S\\
uvwz C_+ & i uvz^2 S & i v^2wz S & v^2z^2 C_+
\end{pmatrix},
\label{eq:rho_phi_eff_matrix}
\end{equation}
where
\begin{equation}
C_\pm:=2(1\pm\cos\varphi),\qquad S:=2\sin\varphi.
\end{equation}

The incoherent which-branch mixture
\begin{equation}
\rho_{\rm inc}
=
\frac12
\left(
\ket{\Gamma_1}\bra{\Gamma_1}
+
\ket{\Gamma_2}\bra{\Gamma_2}
\right)
\end{equation}
takes the form
\begin{equation}
\rho_{\rm inc}
=
\begin{pmatrix}
u^2w^2 & 0 & 0 & uvwz\\
0 & u^2z^2 & uvwz & 0\\
0 & uvwz & v^2w^2 & 0\\
uvwz & 0 & 0 & v^2z^2
\end{pmatrix}.
\label{eq:rho_inc_eff_matrix}
\end{equation}
Therefore the dephased encoded Bell state
\begin{equation}
\rho_{p,\varphi}
=
(1-p)\rho_\varphi+p\rho_{\rm inc},
\qquad 0\le p\le 1,
\end{equation}
is obtained exactly from Eqs.~\eqref{eq:rho_phi_eff_matrix} and
\eqref{eq:rho_inc_eff_matrix}.

In this product basis, the partial transpose on subsystem \(B\) is the transpose of
each \(2\times2\) block:
\begin{equation}
\rho_{p,\varphi}^{T_B}
=
\begin{pmatrix}
A^T & B^T\\
C^T & D^T
\end{pmatrix}
\quad\text{if}\quad
\rho_{p,\varphi}
=
\begin{pmatrix}
A & B\\
C & D
\end{pmatrix},
\label{eq:PT_block_transpose}
\end{equation}
where \(A,B,C,D\) are \(2\times2\) blocks. The negativity is
\begin{equation}
\mathcal N(\rho_{p,\varphi})
=
\frac{\|\rho_{p,\varphi}^{T_B}\|_1-1}{2}
=
\sum_{\lambda_i<0}|\lambda_i|,
\label{eq:negativity_from_PT_spectrum}
\end{equation}
where \(\{\lambda_i\}\) are the eigenvalues of \(\rho_{p,\varphi}^{T_B}\).

For the in-phase case \(\varphi=0\), define
\begin{equation}
s:=\sqrt{(1-a^2)(1-b^2)},\qquad
\Delta:=2-p(1-ab).
\end{equation}
Then \(\mathcal Z_0=2(1+ab)\), and \(\rho_{p,0}^{T_B}\) is block diagonal up to a
basis permutation. Its four eigenvalues are
\begin{align}
\lambda_{1,2}
&=
\frac{1}{2}
\left[
1-\frac{p}{2}(1-ab)
\pm
\frac{1}{2}
\sqrt{
\frac{(a+b)^2\Delta^2}{(1+ab)^2}
+p^2s^2
}
\right],
\label{eq:PT_eigs_block1_phi0}
\\[4pt]
\lambda_{3,4}
&=
\frac{1}{2}
\left[
\frac{p}{2}(1-ab)
\pm
\frac{1}{2}
\sqrt{
p^2(a-b)^2
+
\frac{s^2\Delta^2}{(1+ab)^2}
}
\right].
\label{eq:PT_eigs_block2_phi0}
\end{align}
For this family there is at most one negative eigenvalue, namely \(\lambda_4\).
Hence
\begin{equation}
\mathcal N(\rho_{p,0})
=
\max\{0,-\lambda_4\}
=
\max\left\{
0,\,
\frac14
\left[
\sqrt{
p^2(a-b)^2
+
\frac{s^2\Delta^2}{(1+ab)^2}
}
-
p(1-ab)
\right]
\right\}.
\label{eq:negativity_closed_phi0}
\end{equation}
For \(p=0\), this reduces to the pure-state result
\[
\mathcal N(\Psi_0)
=
\frac{\sqrt{(1-a^2)(1-b^2)}}{2(1+ab)}.
\]
For \(p=1\), the negativity vanishes, as expected for the fully incoherent
which-branch mixture.

\paragraph{Schmidt decomposition}
\label{app:schmidt_nonorthogonal}

The finite-support construction also gives the Schmidt decomposition of pure
bipartite branch states. Consider two nonorthogonal branch families
\(\{\ket{\psi_i^{A}}\}_{i=1}^{D_A}\) and
\(\{\ket{\psi_j^{B}}\}_{j=1}^{D_B}\), with Gram matrices \(G_A\) and \(G_B\).
A general pure state supported on their product span can be written as
\begin{equation}
\ket{\psi_{AB}}
=
\sum_{i,j} M_{ij}\,
\ket{\psi_i^{A}}\otimes\ket{\psi_j^{B}} .
\label{eq:schmidt_nonorth_state}
\end{equation}
Let
\begin{equation}
\Phi_A=\Psi_A G_A^{-1/2},
\qquad
\Phi_B=\Psi_B G_B^{-1/2}
\end{equation}
be the local L\"owdin orthonormal bases, so that
\(\Psi_A=\Phi_A G_A^{1/2}\) and
\(\Psi_B=\Phi_B G_B^{1/2}\). In these orthonormal bases the same state reads
\begin{equation}
\ket{\psi_{AB}}
=
\sum_{\ell,m}\widetilde M_{\ell m}\,
\ket{\phi_\ell^{A}}\otimes\ket{\phi_m^{B}},
\label{eq:schmidt_orth_basis_state}
\end{equation}
with
\begin{equation}
\widetilde M
=
G_A^{1/2}\,M\,(G_B^{1/2})^{\mathsf T}.
\label{eq:Mtilde_nonorthogonal}
\end{equation}
If the local Gram matrices are real symmetric, as in the two-branch examples
after fixing phases, the transpose in Eq.~\eqref{eq:Mtilde_nonorthogonal} can
be omitted.

The reduced density operators are therefore
\begin{equation}
\rho_A
=
\Phi_A\,
\big(\widetilde M\widetilde M^\dagger\big)\,
\Phi_A^\dagger,
\qquad
\rho_B
=
\Phi_B\,
\big(\widetilde M^{\mathsf T}\widetilde M^*\big)\,
\Phi_B^\dagger .
\label{eq:reduced_states_from_Mtilde}
\end{equation}
Thus the nonzero eigenvalues of
\(\widetilde M\widetilde M^\dagger\) and
\(\widetilde M^\dagger\widetilde M\) are the Schmidt weights of
\(\ket{\psi_{AB}}\). Equivalently, if
\begin{equation}
\widetilde M
=
U\,\mathrm{diag}(s_r)\,V^\dagger
\end{equation}
is the singular-value decomposition, then
\begin{equation}
\ket{\psi_{AB}}
=
\sum_r s_r\,\ket{u_r^A}\otimes\ket{v_r^B},
\qquad
\sum_r s_r^2=1,
\label{eq:schmidt_svd}
\end{equation}
where \(s_r^2\) are the eigenvalues of the reduced density matrices. Explicitly,
$\ket{u_r^A}=\Phi_A U e_r$ and $\ket{v_r^B}=\Phi_B V^*e_r$, where
$e_r$ is the $r$-th coordinate vector. The complex conjugation in the
$B$-side Schmidt vectors follows from the coefficient convention in
Eq.~\eqref{eq:schmidt_orth_basis_state}.

\section{Appendix E: Entanglement entropy of Gaussian-branch GHZ states}
\label{app:ghz_entropy_scaling}

Consider \(N\) bosonic modes in a superposition of two product Gaussian
configurations,
\begin{equation}
\ket{\Psi_N}
=\frac{1}{\sqrt{\mathcal Z_N}}
\left(\ket{g_+}^{\otimes N}+\ket{g_-}^{\otimes N}\right),
\qquad
\mathcal Z_N=2(1+s^N),
\label{eq:ghz_branch_state_app}
\end{equation}
where \(s:=\langle g_+|g_-\rangle\) and the phases of \(\ket{g_\pm}\) are chosen
so that \(s\ge 0\). This is the Gaussian-branch analogue of a GHZ state, and it
contains multipartite entangled coherent states~\cite{WangSanders2002,Sanders2012}
and cat-like bosonic resources~\cite{Brune1996Cat,Vlastakis2013Cat,Haroche2013RMP}
as special cases. Unlike in a qubit GHZ state, the two local branches are not
orthogonal, so the entanglement across a cut depends on how well each side can
tell the two branches apart.

Split the modes into a block \(A\) of \(\ell\) modes and its complement
\(\bar A\) of \(N-\ell\) modes, and write
\(\ket{A_\pm}:=\ket{g_\pm}^{\otimes\ell}\) and
\(\ket{B_\pm}:=\ket{g_\pm}^{\otimes(N-\ell)}\).
Across this cut, \(\ket{\Psi_N}\) is a two-branch product superposition of the
form of Eq.~\eqref{eq:bell}, with collective overlaps
\begin{equation}
x:=\langle A_+|A_-\rangle=s^\ell,
\qquad
y:=\langle B_+|B_-\rangle=s^{N-\ell}.
\label{eq:collective_ghz_overlaps}
\end{equation}
Its Schmidt rank is therefore at most two for any number of modes and any
photon number, and its spectrum follows from Eqs.~\eqref{eq:negativity}--\eqref{eq:schmidt}
with \(a=x\), \(b=y\) and \(\varphi=0\). Explicitly, in the even and odd
combinations
\begin{equation}
\ket{e^A_\pm}=\frac{\ket{A_+}\pm\ket{A_-}}{\sqrt{2(1\pm x)}},
\qquad
\ket{e^B_\pm}=\frac{\ket{B_+}\pm\ket{B_-}}{\sqrt{2(1\pm y)}},
\end{equation}
the state reads
\begin{equation}
\ket{\Psi_N}
=\sqrt{\frac{(1+x)(1+y)}{2(1+xy)}}\,\ket{e^A_+}\ket{e^B_+}
+\sqrt{\frac{(1-x)(1-y)}{2(1+xy)}}\,\ket{e^A_-}\ket{e^B_-},
\end{equation}
so the two nonzero eigenvalues of
\(\rho_A=\mathrm{Tr}_{\bar A}\ket{\Psi_N}\!\bra{\Psi_N}\) are
\begin{equation}
\lambda_\pm(\ell,N)
=\frac{(1\pm s^\ell)(1\pm s^{N-\ell})}{2(1+s^N)},
\label{eq:ghz_branch_schmidt_values}
\end{equation}
and the entanglement entropy is
\begin{equation}
S_A(\ell,N)
=-\sum_{\sigma=\pm}\lambda_\sigma(\ell,N)\log_2\lambda_\sigma(\ell,N).
\label{eq:ghz_branch_entropy}
\end{equation}

For orthogonal branches (\(s=0\)) every nontrivial cut carries exactly one bit,
as for a qubit GHZ state. For \(s>0\) the entropy depends on the cut through
\(s^\ell\) and \(s^{N-\ell}\). A small block cannot resolve the two branches and
carries little entanglement; the one-bit value is approached only when both
\(s^\ell\) and \(s^{N-\ell}\) are small, that is, when both sides of the cut can
distinguish the branches. The entropy is symmetric under \(\ell\leftrightarrow N-\ell\).

For displaced coherent branches,
\begin{equation}
\ket{g_\pm}=\ket{\beta\pm\delta},
\label{eq:bright_displaced_branches}
\end{equation}
with real \(\beta\) and \(\delta\), the overlap is
\begin{equation}
s=\langle\beta+\delta|\beta-\delta\rangle=e^{-2\delta^2}.
\label{eq:bright_branch_overlap}
\end{equation}
The entropy thus depends on the separation \(\delta\) but not on the common
displacement \(\beta\): increasing \(\beta\) raises the mean photon number per
mode, \(\bar n=\beta^2+\delta^2\), without changing the entanglement.

\begin{figure}[t]
    \centering
    \includegraphics[width=0.8\linewidth]{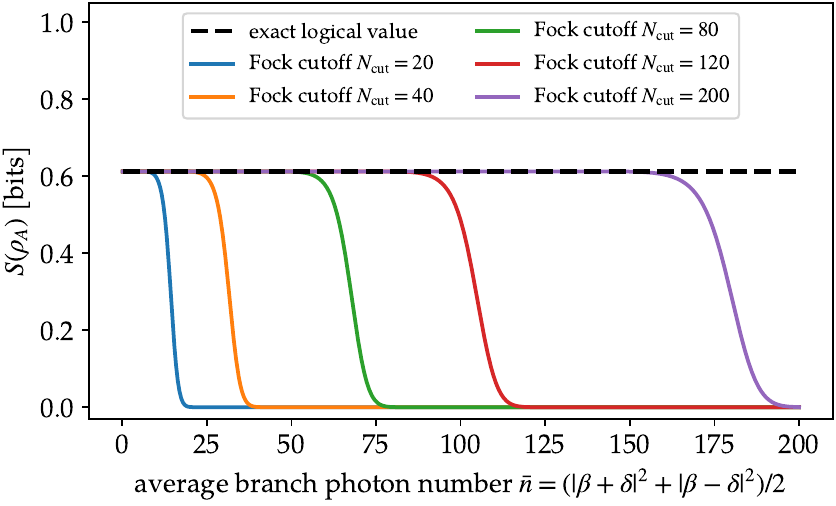}
    \caption{Entanglement entropy of the \(N\)-mode Gaussian-branch GHZ state
    \(\ket{\Psi_N}\propto\ket{\beta+\delta}^{\otimes N}+\ket{\beta-\delta}^{\otimes N}\)
    for \(N=10\), a cut into \(\ell=5\) and \(N-\ell=5\) modes, and \(\delta=0.3\),
    as a function of the mean photon number per mode \(\bar n=\beta^2+\delta^2\).
    Dashed line: exact value from Eq.~\eqref{eq:ghz_branch_entropy} with
    \(s=e^{-2\delta^2}\), independent of \(\beta\). Solid lines: entropy of the
    state truncated at \(N_{\rm cut}\) photons per mode.}
    \label{fig:ghz_entropy_scaling}
\end{figure}

A calculation in a truncated Fock basis does not share this property.
Figure~\ref{fig:ghz_entropy_scaling} compares Eq.~\eqref{eq:ghz_branch_entropy}
with the entropy obtained after truncating every mode at \(N_{\rm cut}\) photons,
for \(N=10\), \(\ell=5\) and \(\delta=0.3\) (\(s\simeq0.835\)). Because the cutoff
acts mode by mode, the truncated state \(P^{\otimes N}\ket{\Psi_N}\), with \(P\)
the projector onto the first \(N_{\rm cut}+1\) Fock states, is again a
superposition of two product states. Its entropy therefore follows from the
truncated single-mode norms \(\langle g_\pm|P|g_\pm\rangle\) and overlap
\(\langle g_+|P|g_-\rangle\), and coincides with a direct diagonalization in the
truncated Fock space, which would involve \((N_{\rm cut}+1)^N\approx10^{23}\)
amplitudes for \(N_{\rm cut}=200\). Each truncated estimate agrees with the exact
value at low \(\bar n\) and drops to zero as \(\bar n\) approaches \(N_{\rm cut}\):
the cutoff removes more weight from the brighter branch \(\ket{\beta+\delta}\),
and the truncated state approaches a single product branch.

\end{document}